\documentclass[]{piparticle-final}
\usepackage{graphicx}
\usepackage{amsmath}
\usepackage{cite} % To compress citation ranges
\usepackage{epstopdf}

\begin{document}

\volume{4}               % To be inserted by Editor
\articlenumber{040002}   % To be inserted by Editor
\journalyear{2012}       % To be inserted by Editor
\editor{J-P. Hulin}   % To be inserted by Editor
%\reviewers{Reviewer's name}  % To be inserted by Editor
\received{29 August 2011}     % To be inserted by Editor
\accepted{29 February 2012}   % To be inserted by Editor
\runningauthor{R Gonz\'alez \itshape{et al.}}  % To be inserted by Editor
\doi{040002}         % To be inserted by Editor

\title{Beltrami flow structure in a diffuser. Quasi-cylindrical approximation }

% Institution references with \cite are inserted after \maketitle in theaffiliation enviroment
\author{Rafael Gonz\'alez,\cite{inst1,inst3}\thanks{E-mail: rgonzale@ungs.edu.ar}\hspace{0.5em}  
        Ricardo Page,\cite{inst2} 
        Andr\'{e}s S. Sartarelli\cite{inst1}}

\pipabstract{We determine the flow structure in an axisymmetric diffuser or expansion region connecting two cylindrical pipes  when the inlet flow is a solid body rotation with a uniform axial flow of speeds $\Omega$ and $U$, respectively. A quasi-cylindrical approximation is made in order to solve the steady Euler equation, mainly the Bragg--Hawthorne equation. As in our previous work on the cylindrical region downstream [R Gonz\'alez et al., Phys. Fluids {\bf 20}, 24106 (2008); R. Gonz\'alez et al., Phys. Fluids {\bf 22}, 74102 (2010), R Gonz\'alez et al., J. Phys.: Conf. Ser. {\bf 296}, 012024 (2011)], the steady flow in the transition region shows a Beltrami flow structure. The Beltrami flow is defined as a field $\mathbf{v}_{B}$ that satisfies $\mbox{\boldmath${\omega}$}_{B}=\nabla \times \mathbf{v}_{B}= \gamma \mathbf{v}_{B}$,  with $\gamma=constant $. We say that the flow has a {\it Beltrami flow structure} when it can be put in the form ${\mathbf v} = U {\mathbf e}_{z}+ \Omega r {\mathbf e}_{\theta}  +{\mathbf v}_{B}$, being $U$ and $\Omega$ constants, i.e it is the superposition of a solid body rotation and translation with a Beltrami one. Therefore, those findings about flow stability hold. The quasi-cylindrical solutions do not branch off and the results do not depend on the chosen transition profile in view of the boundary conditions considered. By comparing this with our earliest work, we relate the critical Rossby number  $\vartheta_{cs}$ (stagnation) to the corresponding one at the fold  $\vartheta_{cf}$ [J. D. Buntine et al., Proc. R. Soc. Lond. A \textbf{449}, 139 (1995)].}

\maketitle

\blfootnote{
\begin{theaffiliation}{99}
   \institution{inst1} Instituto de Desarrollo Humano, Universidad Nacional de General Sarmiento, Gutierrez 1150, 1613 Los Polvorines, Pcia de Buenos Aires, Argentina.
   \institution{inst3} Departamento de F\'\i sica FCEyN, Universidad de Buenos Aires, Pabell\'{o}n I, Ciudad Universitaria, 1428  Buenos Aires, Argentina .
   \institution{inst2} Instituto de Ciencias, Universidad Nacional de General Sarmiento, Gutierrez 1150, 1613 Los Polvorines, Pcia de Buenos Aires, Argentina.
\end{theaffiliation}
}

\section{Introduction}
We have recently  conducted studies on the formation of Kelvin waves and some of their features when an axisymmetric Rankine flow experiences a soft expansion between two cylindrical pipes \cite{Go08,Go10}. One of the significant characteristics of this phenomenon is that the downstream flow shows a Rankine flow superposing a Beltrami flow (Beltrami flow structure \cite{footnote1})). Yet, upstream and downstream cylindrical geometries were considered without taking into account the flow in the expansion. 
This work considered that the base upstream flow, formed by a vortex core surrounded by a potential flow, would have the same Beltrami structure at the expansion and downstream. Nevertheless, the flow at the expansion was not determined. However, it has been seen that this flow is only possible when no reversed flow is present and if its parameters do not take the values where a vortex breakdown appears \cite{Lopez,Benjamin, Guarga}. 
The starting point in the study of the expansion flow is an axysimmetric steady state resulting from the Bragg--Hawthorne equation \cite{Benjamin,Bragg,Batchelor,Alek} for both the vortex breakdown and the formation of waves. Therefore, the solution behavior, whether it branches off or shows a possible stagnation point on the axis, will be determinant to delimit both phenomena. 

Our previous research focused on the formation of Kelvin waves with a Beltrami flow structure downstream \cite{Go08,Go10,Go11}, when the upstream flow was a Rankine one. This present investigation considers only a solid body rotation flow with uniform axial flow at the inlet. As a first step in the study of the flow at the expansion, we only study the rotational flow. However, comparisons with our previous work \cite{Go08} will be drawn.

The aim of this present work is to obtain the steady flow structure at the expansion, considering a quasi-cylindrical approximation when the inlet flow is a solid body rotation with uniform axial flow of speeds $\Omega$ and $U$, respectively. If $a$ is the radius of the cylindrical region upstream, a relevant parameter is the Rossby number $\vartheta=\frac{U}{\Omega a}$. Thus, we would like to determine how this flow depends on the Rossby number, on the geometrical parameters of the expansion and on the critical values of the parameters. We focus on finding the parameter values for which a stagnation point emerges on the axis, or for which the solution of the Bragg--Hawthorne equation branches off. We take them as the conditions for the vortex breakdown to develop.

First, this  paper presents the inlet flow and the corresponding Bragg--Hawthorne equation written for the transition together with the boundary conditions in section II. Second, it works on the quasi-cylindrical approximation for the Bragg--Hawthorne equation and its solution is developed in section III. Third, results and discussions are offered in section IV together with a comparison with our previous work \cite{Go08}.  Finally, conclusions are presented in section V.

\section{The Bragg--Hawthorne equation}
\label{sec:BHE}
We assume an upstream flow in a pipe of radius $a$ as an inlet flow in an axisymmetric expansion of length $L$ connecting to another pipe with radius $b$, $b > a$. The inlet flow filling the pipe consists of a solid body rotation of speed  $\Omega$ with a uniform axial flow of speed  $U$:

\begin{align}
{\mathbf v} = U {\mathbf e}_{z}+ \Omega r {\mathbf e}_{\theta},
\label{rank}
\end{align}
$U$ and $\Omega$ being constants.
The equilibrium flow in the whole region is determined by the steady Euler equation which can be written as the Bragg--Hawthorne equation \cite{Batchelor} 

\begin{align}
\frac{\partial^{2} \psi}{\partial z^{2}}+r \frac{\partial}{\partial r} \left( \frac{1}{r}\frac{\partial \psi}{\partial r}\right)+r^2 \frac{\partial H}{\partial \psi}+ C \frac{\partial C}{\partial \psi}=0,
\label{bhe}
\end{align}
where $\psi$ is the defined stream function 

\begin{align}
v_{r}=-\frac{1}{r}\frac{\partial \psi}{\partial z},\  v_{z}=\frac{1}{r}\frac{\partial \psi}{\partial r},
\label{funco}
\end{align}
and $H(\psi), C(\psi)$ are the total head and the circulation, respectively

\begin{align}
H(\psi)=\frac{1}{2}(v_{r}^{2}+v_{\theta}^{2}+v_{z}^{2})+\frac{p}{\rho},\ C(\psi)=r v_{\theta}.
\label{invariants1}
\end{align}

To solve Eq. (\ref{bhe}), the boundary conditions must be established. These consist of giving the inlet flow, of being both the centerline and the boundary wall, streamlines, and of being the axial velocity positive ($v_{z}>0$). For the upstream flow, the stream function is $\psi=\frac{1}{2}U {r}^{2}$, and $H(\psi), C(\psi)$ are given by

\begin{align}
H(\psi)=\frac{1}{2}U^{2}+\Omega \gamma \psi,\ C(\psi)=\gamma \psi,
\label{invariants2}
\end{align}
$\gamma=\frac{2U}{\Omega}$ being the eigenvalue of the flow with Beltrami structure \cite{Go11}. Thus, by considering the inlet flow, Eqs. (\ref{invariants2}) are valid for the whole region. The second condition regarding the streamlines implies the following relations

\begin{align}
&\psi (r=0, z)=0, \notag \\
&\psi (r=\sigma (z), z)=\frac{1}{2}U {a}^{2}, \ 0\leq z \leq L     
\label{boundaries}
\end{align}
where $r=\sigma (z)$ gives the axisymmetric profile of the pipe expansion. Deducing from Eq.  (\ref{boundaries}), the boundary conditions are determined by the inlet flow.  Additionally, curved profiles are considered, so

\begin{align}
\frac{\partial \psi}{\partial z}(r, z=L)=0, \ 0\leq r \leq b .    
\label{otherboundary}
\end{align}

\section{Quasi-cylindrical approximation}
\label{sec:qca}
If we  consider that $\frac{\partial^{2} \psi}{\partial z^{2}}=0$, the solutions to Eqs. (\ref{bhe}) and (\ref{invariants2}) for the cylindrical regions are given by \cite{Batchelor}

\begin{align}
\psi= \frac{1}{2}U {r}^{2}+ArJ_{1}[\gamma r],
\label{solcil}
\end{align}
where $A$ is a constant. The quasi-cylindrical approximation consists of taking the dependence of $A(z)$ on $z$ but with the condition $\frac{\partial^{2} \psi}{\partial z^{2}}\approx 0$ compared with the remaining terms of (\ref{bhe}). The amplitude $A(z)$ is then obtained by imposing the boundary conditions (\ref{boundaries}) which depend on the wall profile $r=\sigma (z)$, giving

\begin{align}
A(z)= \frac{1}{2} \frac{U \left(a^{2}-\sigma^{2} (z)\right)}{\sigma (z) J_{1}[\gamma \sigma (z)]}.
\label{amp}
\end{align}

By using the dimensionless quantities $\tilde{r}= \frac{r}{a}$, $\tilde{z}= \frac{z}{a}$, $\tilde{v}= \frac{v}{U}$ the stream function in the  quasi-cylindrical approximation can be written as

\begin{align}
&\tilde{\psi}= \frac{1}{2} {\tilde{r}}^{2}+\tilde{A}(\tilde{z})\tilde{r}J_{1}[\frac{2}{\vartheta} \tilde{r}], \notag \\
&\tilde{A}(\tilde{z})=\frac{1}{2} \frac{ \left(1-\tilde{\sigma}^{2} (\tilde{z})\right)}{\tilde{\sigma} (\tilde{z}) J_{1}[\frac{2}{\vartheta} \tilde{\sigma} (\tilde{z})]},
\label{solqcil}
\end{align}
where $\vartheta=\frac{U}{\Omega a}$ is the Rossby number. Hence the velocity field becomes

\begin{eqnarray}
\tilde{v}_{r}(\tilde{r},\tilde{z}) &=&  -\tilde{A}^{'}(\tilde{z}) J_{1}[\frac{2}{\vartheta} \tilde{r}]  \label{solbelt1}\\
&& \nonumber  \\ 
\tilde{v}_{\theta}(\tilde{r},\tilde{z}) &=& \frac{1}{\vartheta} \tilde{r} + \frac{2}{\vartheta} \tilde{A}(\tilde{z}) J_{1}[\frac{2}{\vartheta} \tilde{r}]  \label{solbelt2} \\
&& \nonumber  \\
\tilde{v}_{z}(\tilde{r},\tilde{z}) &=& 1+ \frac{2}{\vartheta} \tilde{A}(\tilde{z}) J_{0}[\frac{2}{\vartheta} \tilde{r}],
\label{solbelt3} 
\end{eqnarray}
where $\tilde{A}^{'}(\tilde{z})={d\tilde{A}(\tilde{z})}/{d\tilde{z}}$. 

Finally, it is necessary to give the wall profile  $\tilde{\sigma} (z)$ to completely determine the flow. Two kinds of profiles were seen:
\begin{description}
	\item[i-] conical profile

\begin{align}
&\tilde{\sigma} (\tilde{z})= 1 +  \left(\frac{\eta - 1}{\tilde{L}}\right)\tilde{z}, \notag\\
&0\leq \tilde{z}  \leq \tilde{L} \text{ and } \eta=\frac{b}{a}.
\label{wall1}
\end{align}
\item[ii-] curved profile

\begin{align}
\tilde{\sigma} (\tilde{z})&= \frac{1 + \eta}{2} -  \left( \frac{\eta - 1}{2} \right) \cos \left( \frac{\pi \tilde{z}}{\tilde{L}} \right), \notag\\
&0\leq \tilde{z}  \leq \tilde{L}.
\label{wall2}
\end{align}
\end{description}
The latter meets the boundary condition  (\ref{otherboundary}) as well. Therefore, Eqs. (\ref{solbelt1}-\ref{wall2}) together with the boundary conditions (\ref{boundaries},\ref{otherboundary}) allow  to determine the flow structure for both the conical and curved wall profile.
\section{Results and discussion}
 \label{sec:results}
We note that the flow keeps a Beltrami flow structure in the quasi-cylindrical approximation. Effectively, giving (\ref{solbelt1}-\ref{solbelt3}) 

\begin{eqnarray}
\tilde{v}_{r}(\tilde{r},\tilde{z}) &=&  \tilde{v}_{Br}(\tilde{r},\tilde{z})  \label{solbelt1b}\\
&& \nonumber  \\ 
\tilde{v}_{\theta}(\tilde{r},\tilde{z}) &=& \frac{1}{\vartheta} \tilde{r} + \tilde{v}_{B \theta}(\tilde{r},\tilde{z})  \label{solbelt2b} \\
&& \nonumber  \\
\tilde{v}_{z}(\tilde{r},\tilde{z}) &=& 1+ \tilde{v}_{Bz}(\tilde{r},\tilde{z}),
\label{solbelt3b} 
\end{eqnarray}
it is easy to see that under this approximation $\nabla\times \mathbf{v}_{B}(\tilde{r},\tilde{z})=\frac{2}{\vartheta} \mathbf{v}_{B}(\tilde{r},\tilde{z})$
and so,  the  whole flow is the sum of a solid body rotation flow with a uniform axial flow plus a Beltrami flow, given the latter in a system with uniform translation velocity $\mathbf{U}=1. \mathbf{\hat{z}}$ and uniform rigid rotation velocity $\mathbf{V}=\frac{1}{\vartheta} \tilde{r} \mbox{\boldmath${\hat{\theta}}$}$.

Given the flow field and its structure, the parameters are considered by evaluating the behavior of  $\tilde{v}_{z}(\tilde{r},\tilde{z_{0}})$ with $\tilde{z_{0}}=\tilde{L}$ i.e., taken at outlet, and with $\tilde{L}=1$. In order to do so, a wall profile is selected  (\ref{wall1} or \ref{wall2}) and three different values of the expansion parameter are taken, mainly  $\eta_{1}=1.1,\  \eta_{2}=1.2$ and $\eta_{3}=1.3$. 

\begin{figure*}[th]
\begin{center}
\includegraphics[width=0.8\textwidth]{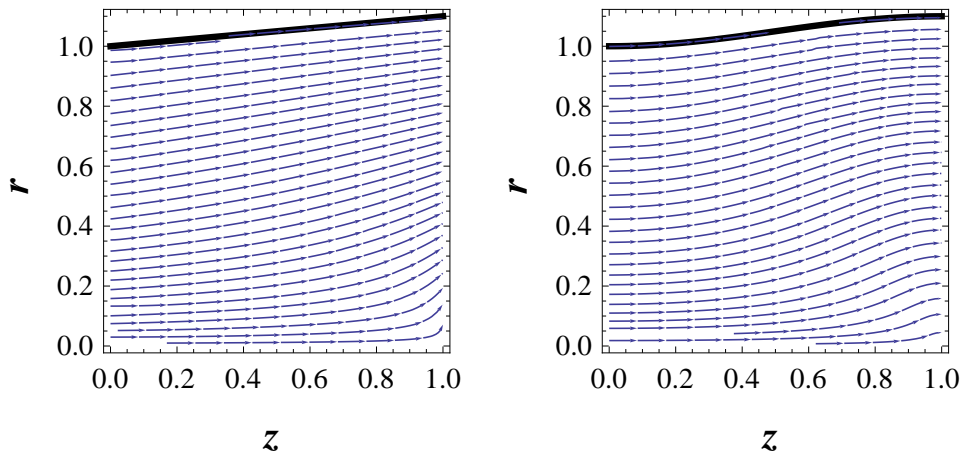}
\caption{Contour flow in the transition region for conical and curved profiles for $\eta_{1}=1.1 $, $\vartheta_{1}=0.695 $.} \label{figure1}
\end{center}
\end{figure*} 

\begin{figure*}[th]
\begin{center}
\includegraphics[width=0.8\textwidth]{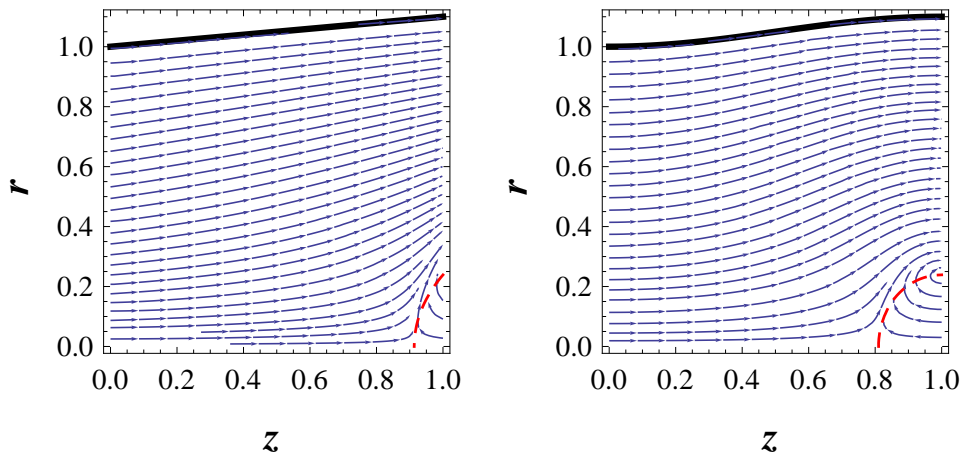}
\caption{Contour flow in the transition region for conical and curved profiles for $\eta_{1}=1.1 $, $\vartheta_{1}=0.68 $. The broken lines represent points with $\tilde{v}_{z}=0$. } \label{figure2}
\end{center}
\end{figure*}

The first step is to analyze the flow dependence on the Rossby number. In Fig. \ref{figure1}, the contour flows corresponding to the conical and curved  profiles for  $\eta_{1}=1.1$, $\vartheta_{1}=0.695 $ are shown. Graphics in Fig. \ref{figure2} represent the same configuration  but for $\vartheta = 0.68 < \vartheta_{1}$. The broken lines represent points for which $\tilde{v}_{z}=0$. Inflow and recirculation are present but it is not a real flow because the model fails when considering inflow. It can be seen that for $\vartheta_{1}=0.695$,  $\tilde{v}_{z}=0$ at the outlet, on the axis. For the Rossby numbers with  $\vartheta \geq \vartheta_{c}$, the azimuthal flow vorticity is negative ($\omega_{\phi} < 0$), resulting in  an increase in the axial velocity with the radius, and so having  a minimum on the axis where the stagnation point appears \cite{Lopez}. Therefore, the critical Rossby number can be defined $\vartheta_{c}$  as the value where  $\tilde{v}_{z}$ is zero at the outlet on the axis i.e., where the flow shows a stagnation point. This is the necessary condition to produce a vortex breakdown  \cite{Lopez}. We find the same critical Rossby number for both wall profiles and so we will not treat them separately from now on. The critical Rosssby values for $\eta_{2}=1.2$ and $\eta_{3}=1.3$ are $\vartheta_{2}=0.869$ and  $\vartheta_{3}=1.052$, respectively.

\begin{figure}[th]
\begin{center}
\includegraphics[width=0.5\textwidth]{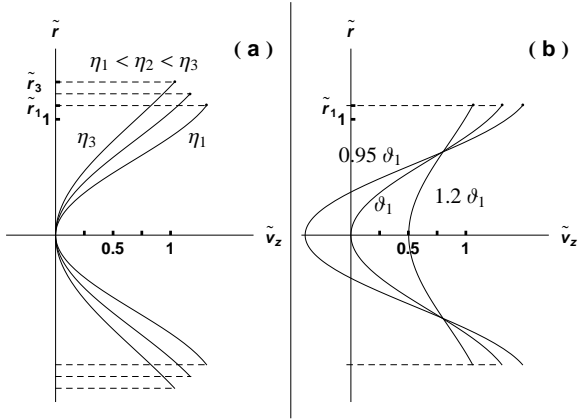}
\caption{(a) $\tilde{v}_{z}$ at the outlet as a function of $r$ for $\eta_{1}$,$\eta_{2}$,$\eta_{3}$ and the corresponding critical Rossby numbers $\vartheta_{1}$,$\vartheta_{2}$,$\vartheta_{3}$. (b) $\tilde{v}_{z}$ at the outlet as a function of $r$ for $\vartheta_{1}$ and for values of $\vartheta$ slightly shifted  from $\vartheta_{1}$ . In each case, the minimum of $\tilde{v}_{z}$ is reached on the axis.} \label{figure3}
\end{center}
\end{figure}

\begin{figure}[th]
\begin{center}
\includegraphics[width=0.4\textwidth]{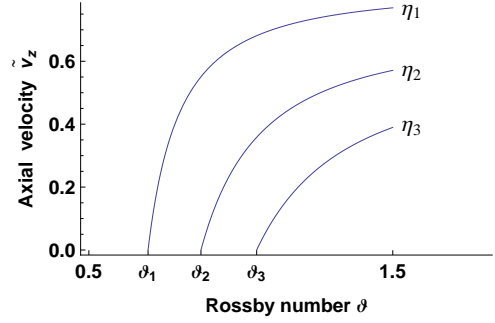}
\caption{$\tilde{v}_{z}$ at the outlet on the axis as a function of the Rossby number $\vartheta$ for $\eta_{1}=1.1,\eta_{2}=1.2$, $\eta_{3}=1.3$. Here $\vartheta_{1}=0.695,\vartheta_{2}=0.869 $ and $\vartheta_{3}=1.052 $ correspond to stagnation points.} \label{figure4}
\end{center}
\end{figure}

Given the previous analysis, the second step is to show the behavior of $\tilde{v}_{z}$ on the axis at the outlet as a function of $\vartheta$ for each $\eta$ in order to study the existence of folds in the Rossby number-continuation parameter (equivalent to the swirl parameter in  \cite{Benjamin,Alek,buntsaff}); indeed, we have seen that $\tilde{v}_{z}$ has the minimum  on the axis. Besides, when using Eq. (\ref{solbelt3}) when $r=0$, it is easy to see that $\tilde{v}_{z}$ decreases with $z$ and so it reaches the minimum at the outlet being $\tilde{v}_{z}\geq 0$. In Fig.~\ref{figure3}, the radial dependence of $\tilde{v}_{z}$ is plotted at the outlet for  $\eta_{1}$,$\eta_{2}$,$\eta_{3}$  and its variation with $\vartheta$ when it is slightly shifted from $\vartheta_{1}$. In Fig. \ref{figure4}, it can be seen that the minimum of $\tilde{v}_{z}$ on the axis increases with $\vartheta$ so there is no fold of $\tilde{v}_{zmin}$ as defined by Buntine and Saffman in a similar approximation \cite{buntsaff}. 

\begin{figure}[th]
\begin{center}
\includegraphics[width=0.4\textwidth]{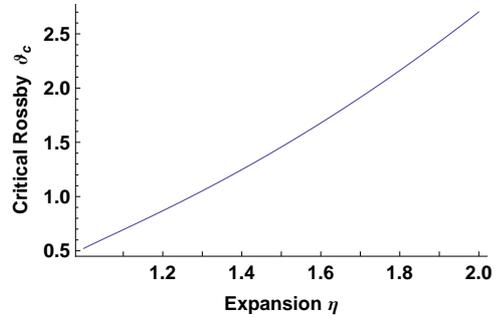}
\caption{Critical Rossby number $\vartheta_{c}$  as a function of $\eta$.} \label{figure5}
\end{center}
\end{figure}

The dependence of the results on $L$ is analyzed. It can be seen that when $z=L$  in Eqs. (\ref{wall1}) and (\ref{wall2}), $\tilde{\sigma} (\tilde{L})=\eta$ is obtained. By replacing this in Eq. (\ref{solbelt3}) for $z=L$ and $r=0$ it gives

\begin{align}
\tilde{v}_{z}min = 1+ \frac{ \left(1-\eta^{2} \right)}{\vartheta \eta J_{1}[\frac{2}{\vartheta} \eta]},
\label{vmin} 
\end{align}
and so $\vartheta_{c}$ is obtained as a function of $\eta$ by solving the last equation when $\tilde{v}_{z}min=0$, as shown in Fig. \ref{figure5}. This result seems to be surprising, but it is not so if it is considered as derived from the quasi-cylindrical approximation: the dependence of the flow on $z$ is obtained through the boundary conditions expressed by Eq. (\ref{boundaries}). At the same time, these boundary conditions depend on the inlet flow and on the parameter $\eta$. This explains the fact that the same results, for both conical and curved profiles, have been obtained and that the condition given by Eq. (\ref{otherboundary}) at the outlet has not influenced them. 

Differences with Batchelor's seminal work should be marked  \cite{Batchelor}. Mainly, he works in cylindrical geometry and does not consider the dependence of the flow on $z$ . We introduce this $z$ dependence through the quasi-cylindrical approximation. This, therefore, allows us to find the structure of the flow in the transition together with the Rossby critical number defined by considering this structure and by showing that the minimum of $\tilde{v}_{z}$ is reached at the outlet on the axis. Nevertheless, once the flow reaches the pipe downstream, the analysis coincides because, as shown, the problem depends on the inlet flow and on the parameter expansion $\eta$. This allows us to consider the issue of the vortex core that we have not considered at the inlet flow. As we know the structure of the flow in the downstream cylindrical region \cite{Go08} and by assuming a quasi-cylindrical approximation for the vortex core in the transition region, the minimum of  ${{{v_{core}}}_{z}}$ at the outlet on the axis is given by

\begin{align}
{{{v_{core}}}_{z}}_{min} = 1+ \frac{ \left(1-\hat{\eta}^{2} \right)}{\hat{\vartheta} \hat{\eta} J_{1}[\frac{2}{\hat{\vartheta}} \hat{\eta}]},
\label{vmincore} 
\end{align}
where $\hat{\vartheta}=\frac{\vartheta}{\iota}$, $\hat{\eta}=\frac{\xi}{\iota}$  and $\xi$ and $\iota$ are the dimensionless radius of the core downstream and upstream, respectively. We note that $\hat{\eta}$ is the expansion parameter of the core. Hence Eqs. (\ref{vmin}) and (\ref{vmincore}) have the same structure. In the present work, we have not found any fold in the Rossby number-continuation parameter of $\tilde{v}_{z}$, as found in our previous work \cite{Go08} where the fold was associated with a critical Rossby number called $\vartheta_{cf}$ by Buntine and Saffman \cite{buntsaff}. As we have already done, we define the Rossby critical number for which ${{{v_{core}}}_{z}}_{min}=0$  where there is a stagnation point, and we will call it ${\vartheta}_{cs}$. In \cite{Go08}, for $\iota=0.272$ and pipe expansion  parameters  $\eta_{1}$,$\eta_{2}$,$\eta_{3}$, we have found that  $\vartheta_{cf}$ were $0.35$, $0.44$ and $0.53$, respectively, while the core expansion  parameters $\hat{\eta}$ were $1.25$, $1.47$ and $1.65$, respectively. 

By replacing these values in Eq. (\ref{vmincore}) when ${{{v_{core}}}_{z}}_{min}$ is zero, we get the corresponding $\hat{\vartheta}_{cs}$ and then ${\vartheta}_{cs}$ for the vortex core. These are respectively $0.26$, $0.38$ and $0.49$. That is to say that in all the cases we have ${\vartheta}_{cs} < \vartheta_{cf}$. Therefore, at the fold $\tilde{v}_{z}>0$. This coincides with the results found by Buntine and Saffman \cite{buntsaff} in their analysis using a three-parameter family inlet flow.

\section{Conclusions}
\label{sec:conclusions} 
The main conclusions drawn from the previous sections are:
\begin{enumerate}
\item In the quasi-cylindrical approximation,  the steady flow in the transition expansion region corresponding to a solid body rotation with uniform axial flow as inlet flow has the same Beltrami flow structure as in the pipe downstream, which is compatible with the boundary conditions. Therefore, findings from our previous work on stability \cite{Go08, Go10, Go11} can hold.
	
\item For fixed values of $\eta$ and $\vartheta \geq \vartheta_{c}$, $\omega_{\phi} < 0$ and then $\tilde{v}_{z}$ in the transition region is an increasing function of $r$ and a decreasing function of $z$ reaching its the minimum on the axis at the outlet.
	
\item For fixed values of $\eta$, the minimum of $\tilde{v}_{z}$ on the axis is an increasing function of $\vartheta$ (Fig. \ref{figure4}), where the stagnation point corresponds to $\vartheta_{c}$.
	
\item As a consequence, no branching off takes place for the solutions of Bragg--Hawthorne equation.

\item The critical Rossby number $\vartheta_{c}$ corresponding to stagnation is an increasing function of $\eta$ (Fig. \ref{figure5}).

\item The whole picture can be reached by putting together these results with those obtained in \cite{Go08}, where there is a branching owing to the boundary conditions at the frontier between the vortex and the irrotational flow. Moreover, since the results in  \cite{Go08} for the rotational flow depend on the inlet flow as well as on the rotational expansion parameter $\hat{\eta}$ defined in Eq. (\ref{vmincore}), given a quasi-cylindrical approximation, it can be concluded that this expression is the minimum of ${v}_{z}$ in the core. Therefore, we can get the critical Rossby number $\vartheta_{cs}$ and compare it with that corresponding to the fold $\vartheta_{cf}$. This present work verifies that  ${\vartheta}_{cs} < \vartheta_{cf}$, in accordance with Buntine and Saffman's results \cite{buntsaff}.

\item In the quasi-cylindrical approximation, previous results do not depend on the chosen profile. This can be explained by the boundary conditions chosen depending on the inlet flow and on the parameter expansion.
\end{enumerate}

\begin{acknowledgements}
We would like to thank Unversidad Nacional de General Sarmiento for its support for this work, and our colleague Gabriela Di Ges\'u for her advice on the English version of this paper. 
\end{acknowledgements}

\end{document}